%% file: prl.tex
\documentclass[reprint, superscriptaddress,nofootinbib]{revtex4-1}

\usepackage[usenames, dvipsnames]{color}
\usepackage{graphicx}

\usepackage{amsmath}
\usepackage{amsfonts}
\usepackage{amssymb}

\usepackage[breaklinks,colorlinks,
   urlcolor=blue,citecolor=blue,linkcolor=blue]{hyperref}
\usepackage[all]{hypcap}

\begin{document}

\newcommand{\pg}{\ensuremath{p\text{-}g}}

\newcommand{\Hnl}{\ensuremath{\mathcal{H}_{pg}}}
\newcommand{\Hgr}{\ensuremath{\mathcal{H}_{!pg}}}
\newcommand{\lnB}{\ensuremath{\ln B^{pg}_{!pg}}}

\newcommand{\REF}[1]{\textcolor{red}{REFERNCE:#1}}

\newcommand{\nevin}[1]{\textcolor{blue}{#1}}
\newcommand{\reed}[1]{\textcolor{BurntOrange}{#1}}
\newcommand{\result}[1]{\textcolor{black}{#1}}


\newcommand{\pressure}{\result{\ensuremath{3.5^{+2.7}_{-1.7}\times10^{34}\ \mathrm{dyn}/\mathrm{cm}^2}}}
\newcommand{\lambdamax}{\result{\ensuremath{730}}}
\newcommand{\lambdamin}{\result{\ensuremath{200}}}


\newcommand{\AupperAbstract}{\result{\ensuremath{\mathrm{few}\times10^{-7}}}}
\newcommand{\AestAbstract}{\result{\ensuremath{10^{-7}}}}

\newcommand{\AupperNinetylogU}{\result{\ensuremath{3.3\times10^{-7}}}}

\newcommand{\AupperNinetyU}{\result{\ensuremath{6.8\times10^{-7}}}}


\newcommand{\FAPlogU}{\result{\ensuremath{50\%}}}


\newcommand{\EnlupperAbstract}{\result{\ensuremath{10^{51}\, \mathrm{ergs}}}}

\newcommand{\EnlupperNinetylogU}{\result{\ensuremath{2.7\times10^{51}\, \mathrm{ergs}}}}


\newcommand{\EgrupperNinetylogU}{\result{\ensuremath{10^{53}\, \mathrm{ergs}}}}


\newcommand{\EnlEgrratio}{\result{\ensuremath{\text{a few percent}}}}


\newcommand{\FestAbstract}{\result{\ensuremath{70\, \mathrm{Hz}}}}


\newcommand{\fgguestimate}{\result{0.5\, \mathrm{Hz}}}


\newcommand{\NmodeupperAbstract}{\result{\mathrm{few}\times10^{2}}}
\newcommand{\Nmodeguestimate}{\result{200}}


\newcommand{\deltaPhihundredlowerNinetylogU}{\result{\ensuremath{7.6\, \mathrm{rad}}}}
\newcommand{\deltaOrbitshundredlowerNinetylogU}{\result{\ensuremath{0.6}}}


\newcommand{\deltaPhithousandlowerNinetylogU}{\result{\ensuremath{32\, \mathrm{rad}}}}
\newcommand{\deltaOrbitsthousandlowerNinetylogU}{\result{\ensuremath{2.6}}}


\newcommand{\lnBAbstract}{\result{0.03^{+0.70}_{-0.58} \text{(maximum a posteriori and 90\% credible region)}}}


\title{
Constraining the $p$-mode--$g$-mode tidal instability with GW170817
}

\include{authors}

\begin{abstract}
We analyze the impact of a proposed tidal instability coupling $p$-modes and $g$-modes within neutron stars on GW170817.
This non-resonant instability transfers energy from the orbit of the binary to internal modes of the stars, accelerating the gravitational-wave driven inspiral.
We model the impact of this instability on the phasing of the gravitational wave signal using three parameters per star: an overall amplitude, a saturation frequency, and a spectral index.
Incorporating these additional parameters, we compute the Bayes Factor (\lnB) comparing our \pg~model to a standard one.
We find that the observed signal is consistent with waveform models that neglect \pg~effects, with $\lnB = \lnBAbstract$.
By injecting simulated signals that do not include \pg~effects and recovering them with the \pg~model, we show that there is a $\simeq \FAPlogU$ probability of obtaining similar \lnB~even when \pg~effects are absent.
We find that the \pg~amplitude for 1.4 $M_\odot$ neutron stars is constrained to $\lesssim \AupperAbstract$, with maxima a posteriori near $\sim \AestAbstract$ and  \pg~saturation frequency $\sim \FestAbstract$.
This suggests that there are less than a few hundred excited modes, assuming they all saturate by wave breaking.
For comparison, theoretical upper bounds suggest a \pg~amplitude $\lesssim 10^{-6}$ and $\lesssim 10^{3}$ modes saturating by wave breaking.
Thus, the measured constraints only rule out extreme values of the \pg~parameters.
They also imply that the instability dissipates $\lesssim \EnlupperAbstract$ over the entire inspiral, i.e., less than $\EnlEgrratio$~of the energy radiated  as gravitational waves.
\end{abstract}

\maketitle

\section{Introduction}
\label{section:introduction}

Detailed analysis of the gravitational-wave (GW) signal received from the first binary neutron star (NS) coalescence event (GW170817 \cite{GW170817}) constrains the tidal deformability of NSs and thus the equation of state (EOS) above nuclear saturation density \cite{GW170817SourceProperties, EOS, De2018}.
Studies of NS tidal deformation typically focus on the linear, quasi-static tidal bulge induced in each NS by its companion.
Such deformations modify the system's binding energy and GW luminosity and thereby alter its orbital dynamics.
The degree of deformation is often expressed in terms of the tidal deformability $\Lambda_i \propto (R_i/m_i)^5$ of each component \cite{Flanagan2008}, or a particular mass-weighted average thereof ($\tilde{\Lambda}$) \cite{GW170817SourceProperties}.
The strong dependence on compactness $R/m$ means that a stiffer EOS, which has larger $R$ for the same $m$, imprints a larger tidal signals than a softer EOS. 
Current analyses of GW data from the LIGO \cite{LIGO} and Virgo \cite{Virgo} detectors favor a soft EOS \cite{EOS, GW170817GRB}.
Specifically, \cite{GW170817SourceProperties} finds $\tilde{\Lambda} \lesssim \lambdamax$ at the 90\% credible level for all waveform models considered, allowing for the components to spin rapidly.
The pressure at twice nuclear saturation density is also constrained to $P = \pressure$ (median and 90\% credible region) \cite{EOS} assuming small component spins.
In addition to GW phasing, the EOS-dependence of $\tilde{\Lambda}$ should correlate with post-merger signals \cite{gw170817postmerger}, possible tidal disruptions, and kilonova observations \cite{gw170817kilonova}.
Observed light-curves for the kilonova suggest a lower bound of $ \tilde{\Lambda}\gtrsim \lambdamin $ \cite{Radice2018, Coughlin2018}.

Although some dynamical tidal effects are incorporated in these analyses (see, e.g., \cite{Hinderer2016, GW170817SourceProperties}), the impact of several types of dynamical tidal effects are neglected because they are assumed to be small or have large theoretical uncertainties. 
These effects arise because tidal fields, in addition to raising a quasi-static bulge, excite stellar normal modes.
Three such excitation mechanisms are (i) resonant linear excitation, (ii) resonant nonlinear excitation, and (iii) non-resonant nonlinear excitation (see, e.g., \cite{Andersson2018}).
The first occurs when the GW frequency (the oscillation frequency of the tidal field) sweeps through a mode's natural frequency (see, e.g., \cite{Lai1994, Reisenegger:94,  Ho1999, Lai:06, Flanagan:07, Yu2017a, Yu2017b, Xu2017}).
However, since the GW frequency increases rapidly during the late inspiral, the time spent near resonance is too short to excite modes to large amplitudes.
As a result, for modes with natural frequencies within the sensitive bands of ground-based GW detectors, the change in orbital phasing is expected to be small ($\Delta \Psi \lesssim 10^{-2}\textrm{ rad}$) unless the stars are rapidly rotating \cite{Ho1999, Lai:06, Flanagan:07}.
The impact of resonant nonlinear mode excitation (i.e., the parametric subharmonic instability) is likewise limited by the swiftness of the inspiral \cite{Weinberg2013}. 

The proposed \pg~tidal instability is a non-resonant, nonlinear instability in which the tidal bulge excites a low-frequency buoyancy-supported $g$-mode and a high-frequency pressure-supported $p$-mode \cite{Weinberg2013, Venumadhav2014, Weinberg2016, Zhou2017}.
It occurs in the inner core of the NS, where the stratification is weak and the shear due to the tidal bulge is especially susceptible to instability.
Unlike resonantly excited modes, an unstable \pg~pair continuously drains energy from the orbit once excited, even after the orbital frequency changes significantly.
There are many potentially unstable \pg~pairs, each becoming unstable at a different frequency and growing at a different rate.
Although there is considerable uncertainty about the number of unstable pairs, their exact growth rates, and how they saturate, estimates suggest that the impact could be measurable with current detectors \cite{Essick2016}.

In this letter, we investigate the possible impact of the \pg~instability on GW170817 using the phenomenological model developed in \cite{Essick2016}.
The model describes the energy dissipated by the instability within each NS, indexed by $i$, in terms of three parameters: (i) an overall amplitude $A_i$, which is related to the number of modes participating in the instability, their growth rates, and their saturation energies, (ii) a frequency $f_i$ corresponding to when the instability saturates, and (iii) a spectral index $n_i$ describing how the saturation energy evolves with frequency. 
In Section \ref{section:phenomenologlical model}, we describe our models in detail.
In Section \ref{section:model selection}, we compare the statistical evidence for models that include the \pg~instability relative to those that do not.
In Section \ref{section:parameter exclusion}, we investigate the constraints on the \pg~parameters from GW170817, and in Section \ref{section:discussion} we conclude.

\section{Phenomenological Model}
\label{section:phenomenologlical model}

Following \cite{Essick2016}, we extend a post-Newtonian (PN) waveform by including a parametrized model of the \pg~instability.
For the initial PN model, we use the \texttt{TaylorF2} frequency-domain approximant (see, e.g., \cite{Buonanno2009}) terminated at the inner-most stable circular orbit, which includes the effects of linear tides ($\tilde{\Lambda}$) and spins aligned with the orbital angular momentum (the impact of mis-aligned spins on \pg~effects is not known).
Waveform systematics between different existing approximants may be important for small \pg~effects.
However, by comparing the waveform mismatches between several other models (\texttt{TaylorF2}, \texttt{SEOBNRT}, \texttt{PhenomDNRT}, and \texttt{PhenomPNRT}, see \cite{GW170817SourceProperties}), we find these systematic become important for \pg~effects roughly an order of magnitude smaller than the upper limits set by our analysis (see Section \ref{section:parameter exclusion}).
We expect \texttt{TaylorF2} to be reasonably accurate and defer a complete analysis of waveform systematics to future work.

Assuming the \pg~effects are a perturbation to \texttt{TaylorF2}, we find that they modify the phase in the frequency-domain by
\begin{widetext}
\begin{equation}
    \Delta \Psi(f) = - \frac{2C_1}{3B^2(3-n_1)(4-n_1)}\left[ \Theta_1 \left(\frac{f}{f_{\rm ref}}\right)^{n_1-3} + (1-\Theta_1) \left(\frac{f_1}{f_{\rm ref}}\right)^{n_1-3}\left( \left(4-n_1\right) - \left(3-n_1\right)\left(\frac{f}{f_1}\right)\right)\right] + (1 \leftrightarrow 2),
    \label{eq:dPsi}
\end{equation}
\end{widetext}
where
$f_i$ is the saturation frequency, $f_\mathrm{ref} \equiv 100\, \mathrm{Hz}$ is a reference frequency with no intrinsic significance, $C_i = [2m_i/(m_1+m_2)]^{2/3} A_i$, $B = (32/5)(G\mathcal{M}\pi f_\mathrm{ref}/c^3)^{5/3}$, $\mathcal{M}=(m_1 m_2)^{3/5}/(m_1+m_2)^{1/5}$, and $\Theta_i = \Theta(f-f_i)$ where $\Theta$ is the Heaviside function.
This approximant is slightly different than  that of \cite{Essick2016} because they incorrectly applied the saddle-point approximation to obtain the frequency-domain waveform from time-domain phasing \cite{Cutler1994}.
This correction renders the \pg~instability slightly more difficult to measure than predicted in \cite{Essick2016}, although the observed behavior is qualitatively similar.
Specifically, we find that in order to achieve the same $|\Delta \Psi|$, $A_i$ needs to be larger than \cite{Essick2016} found by a factor of $\sim (4-n_i)$, although the precise factor also depends on the other \pg~parameters.

The $\Delta \Psi$ expression contains three types of terms: a constant term, a linear term $\propto (1-\Theta_i)f$, and a power-law term $\propto\Theta_i f^{n_i-3}$.
The constant term corresponds to an overall phase offset and is degenerate with the orbital phase at coalescence.
The linear term corresponds to a change in the time of coalescence; because the \pg~instability transfers energy from the orbit to stellar normal modes, the binary inspirals faster than it would if the effect was absent.
The power-law term accounts for the competition between the rate of \pg~energy dissipation and the rate of inspiral, both of which increase as $f$ increases.
As argued in \cite{Essick2016}, we expect $n_i<3$, which implies that the phase shift accumulates primarily at frequencies just above the ``turn-on" (saturation) frequency $f\gtrsim f_i$.  

When $n_i < 3$, \pg~effects are most important at lower frequencies whereas linear tides ($\tilde{\Lambda}$) and spins ($\chi_i = c S_i/G m_i^2$, where $S_i$ is the spin-angular momentum of each component) have their largest impact at higher frequencies (see, e.g., \cite{Agathos2014}).
The priors placed on the latter quantities can, however, affect our inference of \pg~parameters.

In order to account for a possible dependence on the component masses ($m_i$), we parametrize our model using a Taylor expansion in the \pg~parameters around $m_i=1.4M_\odot$ and sample from the posterior using the first two coefficients.
Our model computes $A_i$ as
\begin{equation}
    A_i(m_i) = A_0 + \left(\left.\frac{dA}{dm}\right|_{1.4M_\odot}\right)\left(m_i - 1.4M_\odot\right),
\end{equation}
and uses $A_0$ and $dA/dm$ instead of $A_1$ and $A_2$.
The model uses similar representations for $f_i$ and $n_i$ in terms of the parameters $f_0$, $df/dm$, $n_0$, and $dn/dm$.
We assume a uniform prior on $\log_{10} A_0$ within $10^{-10} \leq A_0 \leq 10^{-5.5}$, a uniform prior in $f_0$ within $10\,\mathrm{Hz} \leq f_0 \leq 100\,\mathrm{Hz}$, and a uniform prior in $n_0$ within $-1 \leq n_0 \leq 3$. 
The priors on the first-order terms ($dA/dm, df/dm, dn/dm$) are the same as those in \cite{Essick2016}; when $m_1 \sim m_2$, they imply $A_1 \sim A_2$, etc.

We investigate GW170817 using data from several different frequency bands and with different spin priors, but unless otherwise noted we focus on results for data above 30 Hz with $|\chi_i|\leq 0.89$.
Throughout this letter, results from GW170817 were obtained using the same data conditioning as \cite{GW170817SourceProperties}, including the removal of a short-duration noise artifact from the Livingston data (\cite{glitch_mitigation} and discussion in \cite{GW170817}) along with other independently measured noise sources (see, e.g., \cite{P1700260, iLIGO_SeisCleaning, Meadors2014, Tiwari2015}), calibration \cite{Cahillane2017, Viets2018}, marginalization over calibration uncertainties, and whitening procedures \cite{Cornish2015, Littenberg2015}.

\section{Model Selection}
\label{section:model selection}

Using GW data from GW170817, we perform Bayesian model selection.
We compare a model that includes linear tides, spin components alinged with the orbital angular momentum, and PN phasing effects up to 3.5 PN phase terms (\Hgr) to an extension of this model that also includes \pg~effects (\Hnl).
Since we have nested models (\Hgr~is obtained from \Hnl~as $A_i\rightarrow0$)\footnote{Since we use  a uniform-in-$\log_{10} A_0$ prior, \Hnl~does not formally include $A_i=0$.  Nonetheless, our lower limit on $A_i$ is sufficiently small that \Hgr~is effectively nested in \Hnl.}, we use the Savage-Dickey Density Ratio (see, e.g., \cite{Dickey1970, Verdinelli1995, Wagenmakers2010}) to estimate the Bayes Factor ($B^{pg}_{!pg}=p(D|\Hnl)/p(D|\Hgr)$, where $D$ refers to the observed data).
Specifically, we sample from the model's posterior distribution \cite{LALInference} and calculate
\begin{widetext}
\begin{align}\label{equation:sddr derivation}
    \lim\limits_{A_i\rightarrow0}\left[\frac{p(A_i|D, \Hnl)}{p(A_i|\Hnl)}\right] & = \lim\limits_{A_i\rightarrow 0}\left[\frac{1}{p(D|\Hnl)} \int d \theta df_i \, dn_i \,  p(D|\theta, A_i, f_i, n_i; \Hnl) \,  p(\theta|\Hnl) \,  p(f_i, n_i|A_i, \Hnl)\right] \nonumber \\
                                                                 & = \frac{1}{p(D|\Hnl)} \int d \theta \, p(D|\theta; \Hgr) \, p(\theta|\Hgr) \left[\frac{p(\theta|\Hnl)}{p(\theta|\Hgr)}\right] \int df_i \, dn_i \,  p(f_i, n_i|A_i, \Hnl) \nonumber \\
                                                                 & = \frac{p(D|\Hgr)}{p(D|\Hnl)} \left< \frac{p(\theta|\Hnl)}{p(\theta|\Hgr)} \right>_{p(\theta|D, \Hgr)},
\end{align}
\end{widetext}
where $\theta$ refers to all parameters besides the \pg~phenomenological parameters, we note that $\int df dn\, p(f_i,n_i|A_i, \Hnl)=1\, \forall\, A_i$, and $\left<x\right>_p$ denotes the average of $x$ with respect to the measure defined by $p$.
Assuming that $p(\theta|\Hnl)=p(\theta|\Hgr)$, we determine \lnB~from the ratio, as $A_i \rightarrow 0$, of the marginal distribution of $A_i$ \textit{a priori} to the distribution \textit{a posteriori}
\begin{equation}\label{equation:sddr}
    \lnB = \lim\limits_{A_i\rightarrow 0} \left[\ln p(A_i|\Hnl) - \ln p(A_i|D, \Hnl)\right].
\end{equation}
We confirmed that this estimate agrees with estimates from both nested sampling \cite{skilling2006} and thermodynamic integration \cite{Nicolas2006}.

\begin{figure}
    \includegraphics[width=\columnwidth]{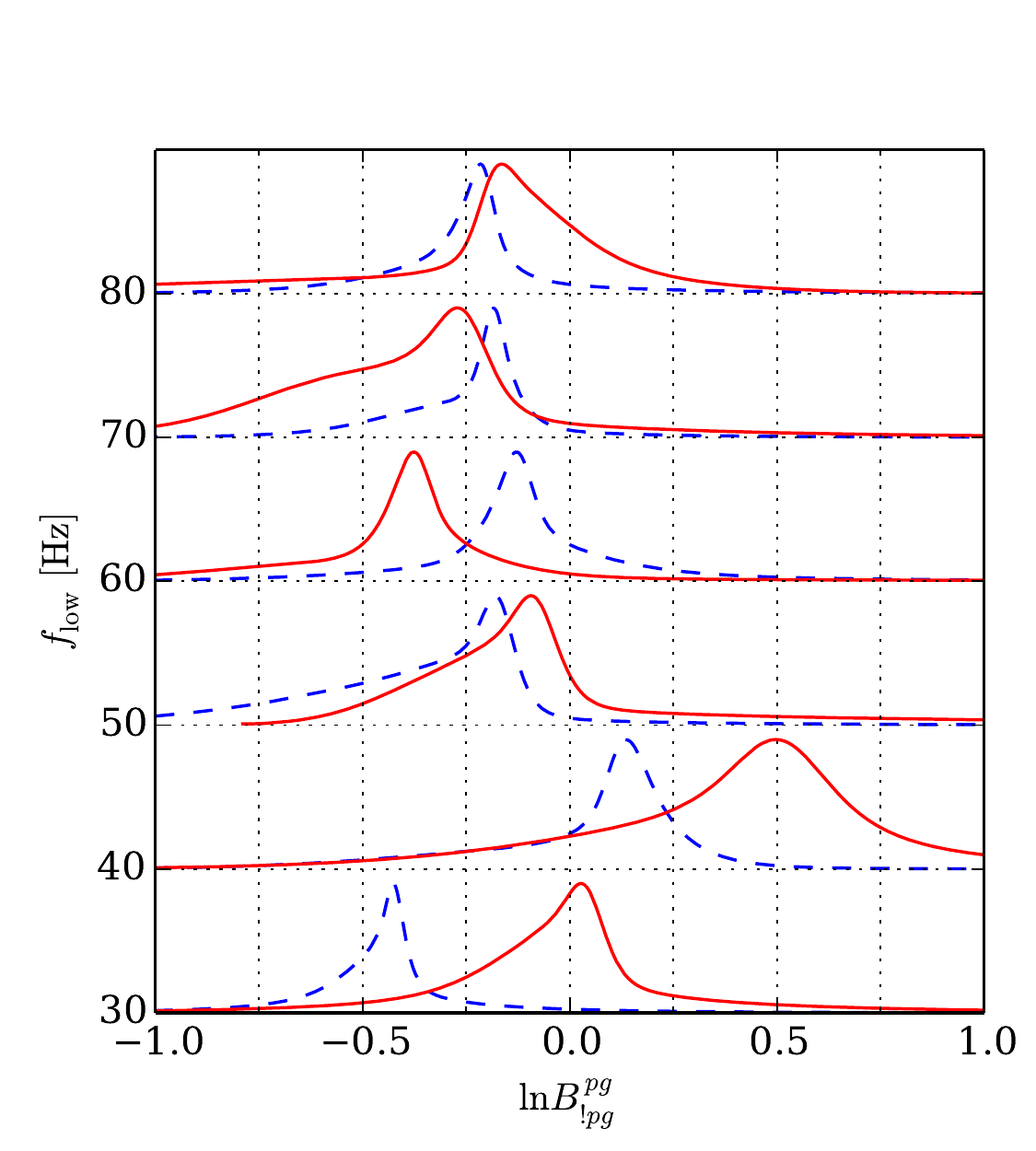}
    \caption{
        Distributions of \lnB~due to sampling uncertainty for different values of $f_\mathrm{low}$.
        The solid red curves assume high-spin priors ($|\chi_i|\leq0.89$) and the dashed blue curves assume low-spin priors ($|\chi_i|\leq0.05$).
    }
    \label{figure:lnB}
\end{figure}

Figure \ref{figure:lnB} shows $\lnB$ as a function of $f_\mathrm{low}$, the minimum GW frequency considered.
At a given $f_{\rm low}$, we show the distribution of $\lnB$ due to the sampling uncertainty from the finite length of our MCMC chains.
The solid and dashed curves correspond to the high-spin ($|\chi_i| \leq 0.89$) and low-spin ($|\chi_i|\leq0.05$) priors discussed in \cite{GW170817,GW170817SourceProperties,EOS}.  

For certain combinations of $f_\mathrm{low}$ and $|\chi_i|$, we find $\lnB>0$, suggesting \Hnl~ is more likely than \Hgr.
In order to assess how likely such values are, we calculate \lnB~ for a large number of simulated, high-spin signals with $A_i=0$ and distinct realizations of detector noise from times near GW170817.
We find that simulated signals without \pg~effects can readily produce $\lnB$ at least as large as the ones we measured from GW170817.
In particular, \lnB~for the 30 Hz, high-spin data corresponds to a False Alarm Probability (FAP) $\approx \FAPlogU$.
We focus on the 30 Hz, high-spin data because it corresponds to the largest bandwidth investigated and the largest signal-to-noise ratio.
The high-spin prior is the most inclusive prior considered, and therefore allows the most model freedom when fitting \pg~effects.

In our model of the instability, the phase shift $\Delta \Psi$ accumulates primarily at frequencies just above the saturation frequency $f\gtrsim f_i$.
Therefore, if it is present, its impact should become more apparent as we decrease the minimum GW frequency considered from $f_\mathrm{low} \gg f_i$ to $f_{\rm low}\lesssim f_i$.
We do see some indication of this behavior in Fig.~\ref{figure:lnB}.
However, we note that if our phenomenological model breaks down at $f<f_i$ due to poor modeling of the pre-saturation behavior (e.g., if our step-function turn-on at $f_i$ is not a good approximation to the instability's induced phase shift), we might expect \lnB~to decrease as we lower $f_\mathrm{low}$ below $f_i$.

\section{Parameter Inference}
\label{section:parameter exclusion}

\begin{figure*}
    \begin{minipage}{0.556875\textwidth}
        \begin{center}
            \includegraphics[width=\textwidth]{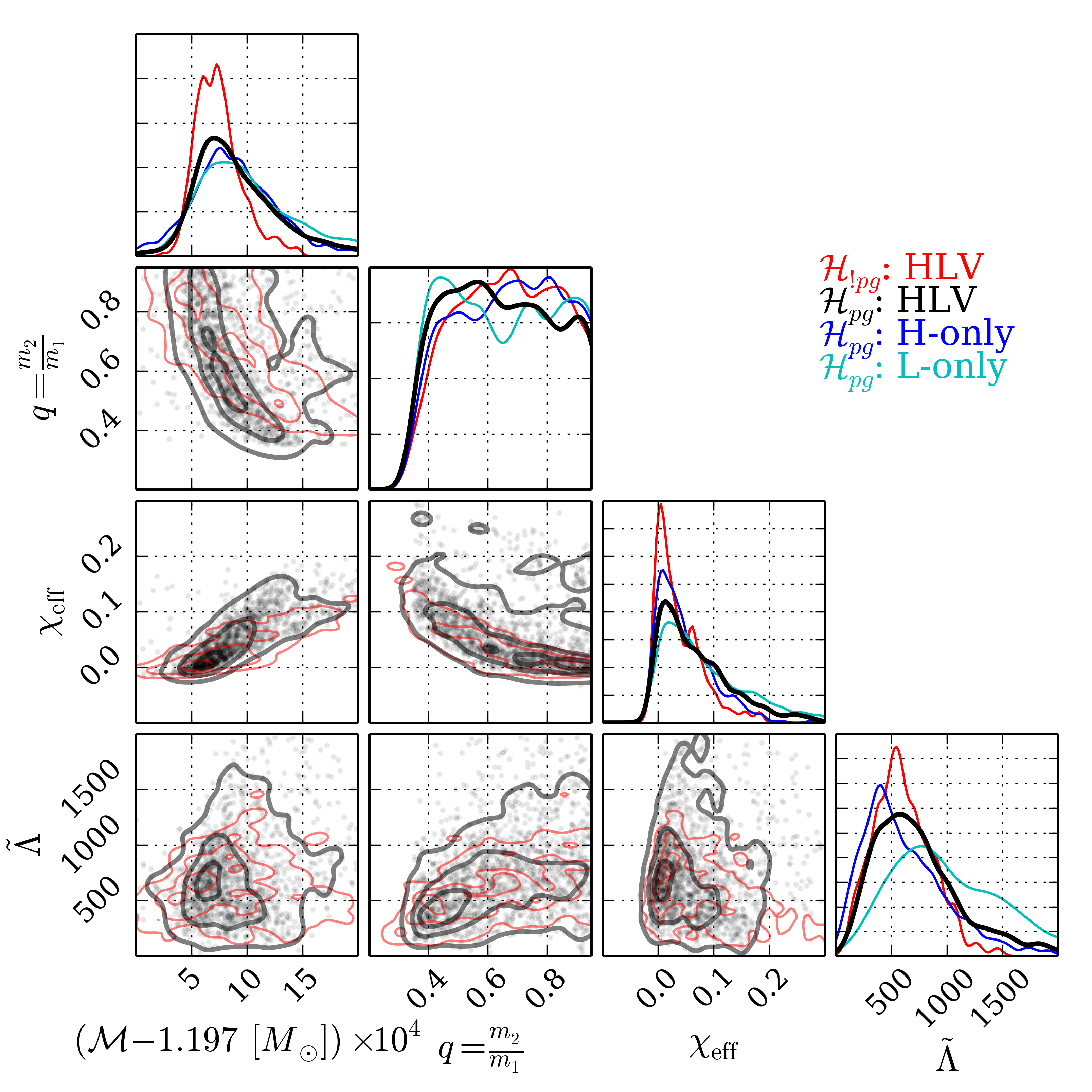} \\
        \end{center}
    \end{minipage}
    \begin{minipage}{0.433125\textwidth}
        \begin{center}
            \vspace{0.5cm}
            \includegraphics[width=\textwidth]{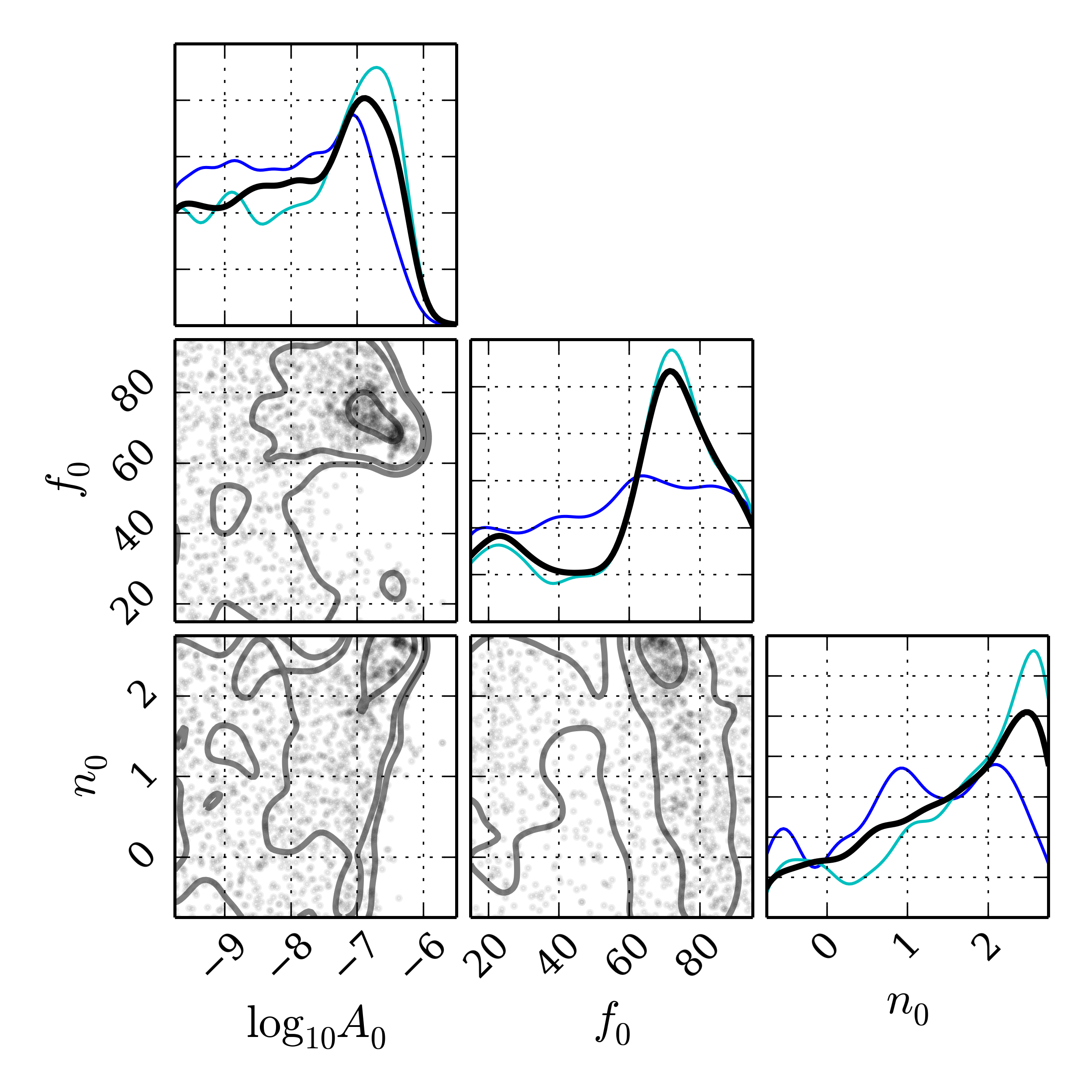} \\
        \end{center}
    \end{minipage}
    \caption{
        Posterior distribributions for \Hgr~(red) and \Hnl with Hanford, Livingston, and Virgo data (thick black, grey shading), Hanford data only (dark blue), and Livingston data only (light blue) using GW data above 30 Hz, $|\chi_i|\leq0.89$, and a uniform-in-$\log_{10} A_0$ prior.
        \textit{Left:} a subset of parameters shared by \Hgr~and \Hnl.
        \textit{Right:} a subset of parameters belonging only to \Hnl.
        We only show one-dimensional posteriors for the single instrument data, although the multi-dimensional posteriors are similarly consistent with the full \Hnl~data.
        Contours in the two-dimensional distributions represent 10\%, 50\%, and 90\% confidence regions for the \Hnl~and \Hgr~models, respectively.
    }
    \label{figure:corner}
\end{figure*}

We now investigate the constraints obtained from GW170817.
Figure \ref{figure:corner} shows the joint posterior distributions for both \Hnl~and \Hgr.
We find that \Hnl~and \Hgr~yield similar posterior distributions for all non-\pg~parameters, including both extrinsic and intrinsic parameters.
The constraints on  the chirp mass ($\mathcal{M}$), effective spin $\chi_\mathrm{eff} = (m_1\chi_1 + m_2\chi_2)/(m_1+m_2)$, and $\tilde{\Lambda}$ are slightly weaker in \Hnl~than \Hgr.
This is because \Hnl~ provides extra freedom to the signal's duration in the time-domain.

Regarding the \pg~parameters, we find a noticeable peak near $A_0\sim10^{-7}$ with a flat tail to small $A_0$.
We find $A_0 \leq \AupperNinetylogU$ assuming a uniform-in-$\log_{10} A_0$ prior and $A_0 \leq \AupperNinetyU$ assuming a uniform-in-$A_0$ prior, both at 90\% confidence.\footnote{The upper limit with a uniform-in-$A_0$ prior is larger only because we weight larger values of $A_0$ more \textit{a priori} than with a uniform-in-$\log_{10} A_0$ prior.}
We also find a peak at $f_0\sim\FestAbstract$.
The peaks persist when we analyze the data from each interferometer separately, with reasonably consistent locations and shapes (Fig.~\ref{figure:corner}).
However, we find that the simulated signals with $A_i=0$ can produce similar peaks, suggesting they may be due to noise alone.
Similar to \cite{Essick2016}, we find that $n_i$ is not strongly constrained and the gradient terms in the Taylor expansions are not measurable.

Theoretical arguments suggest an upper bound of $A_0 \lesssim 10^{-6}$ \cite{Essick2016}.
Therefore, our $A_0$ constraint only rules out the most extreme values of the \pg~parameters.

\section{Discussion}
\label{section:discussion}

While GW170817 is consistent with models that neglect \pg~effects, it is also consistent with a broad range of \pg~parameters.
The constraints from GW170817 imply that there are $\lesssim\Nmodeguestimate$ excited modes at $f=100\,\mathrm{Hz}$, assuming all modes grow as rapidly as possible and saturate at their breaking amplitudes ($\lambda=\beta=1$ in Eq.~(7) of \cite{Essick2016}) and that the frequency at which modes become unstable is well approximated by $f_0$.
For comparison, theoretical arguments suggest an upper bound of $\sim 10^{3}$ modes saturating by wave breaking \cite{Essick2016}. 
More modes may be excited if they grow more slowly or saturate below their wave breaking energy.

\begin{figure}
    \includegraphics[width=\columnwidth]{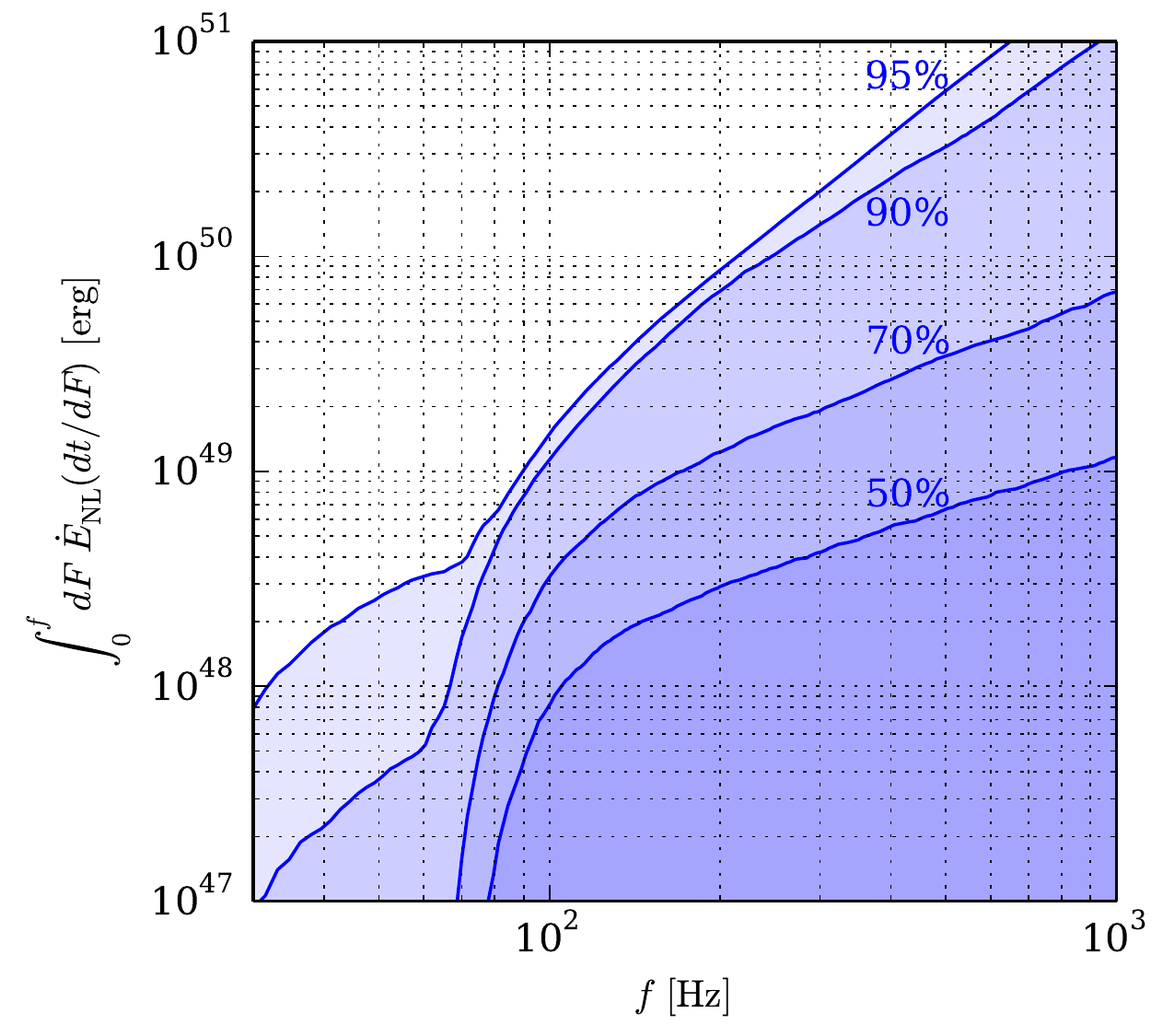}
    \caption{
        Upper limits on the cumulative enegy dissipated by the \pg~instability as a function of frequency.
        We note the relatively strong constraints at lower frequencies, where \pg~effects are more pronounced.
    }
    \label{figure:energy}
\end{figure}

We can also use the measured constraints to place upper limits on the amount of energy dissipated by the \pg~instability.
As Fig.~\ref{figure:energy} shows, \pg~effects dissipate $\lesssim \EnlupperNinetylogU$ throughout the entire inspiral at 90\% confidence.
In comparison, GWs carry away $\gtrsim \EgrupperNinetylogU$.
This implies time-domain phase shifts $|\Delta \phi| \lesssim \deltaPhihundredlowerNinetylogU$ (\deltaOrbitshundredlowerNinetylogU~orbits) at $100\textrm{ Hz}$ and $|\Delta \phi| \lesssim \deltaPhithousandlowerNinetylogU$ (\deltaOrbitsthousandlowerNinetylogU~orbits) at $1000\textrm{ Hz}$ after accounting for the joint uncertainty in component masses, spins, linear tides, and \pg~effects.

A $g$-mode with natural frequency $f_g$ is predicted to become unstable at a frequency $f_{\rm crit}\simeq 45\, \mathrm{Hz} (f_g/10^{-4}\lambda f_{\rm dyn})^{1/2}$, where $f_{\rm dyn}$ is the dynamical frequency of the NS and $\lambda$ is a slowly varying function typically between $0.1-1$ \cite{Weinberg2016,Essick2016}.
Since the modes grow quickly, the frequency at which the instability saturates is likely close to the frequency at which the modes become unstable ($f_0\simeq f_{\rm crit}$).
If we assume that the observed peak near $f_0\sim\FestAbstract$ is not due to noise alone, then the maximum a posteriori estimate for $f_0$ along with approximate values for the masses (1.4 $M_\odot$) and radii (11 km) of the components \cite{EOS} imply $f_g \simeq \fgguestimate$.

With several more signals comparable to GW170817, it should be possible to improve the amplitude constraint to $A_0 \lesssim 10^{-7}$.
Obtaining even tighter constraints will likely require many more detections, especially since most events will have smaller SNR.
Future measurements will also benefit from a better understanding of how the instability saturates.
To date, there have only been detailed theoretical studies of the instability's threshold and growth rate \cite{Weinberg2013, Venumadhav2014, Weinberg2016, Zhou2017}, not its saturation.
As a result, we cannot be certain of the fidelity of our phenomenological model.

While this letter was in final internal review, related work was posted \cite{Reyes2018} with, in particular, the conclusion that the \Hgr~model is strongly favored over the \Hnl~model by a factor of at least $10^4$.
We are investigating possible reasons for the differences between our conclusions.


The authors gratefully acknowledge the support of the United States
National Science Foundation (NSF) for the construction and operation of the
LIGO Laboratory and Advanced LIGO as well as the Science and Technology Facilities Council (STFC) of the
United Kingdom, the Max-Planck-Society (MPS), and the State of
Niedersachsen/Germany for support of the construction of Advanced LIGO
and construction and operation of the GEO600 detector.
Additional support for Advanced LIGO was provided by the Australian Research Council.
The authors gratefully acknowledge the Italian Istituto Nazionale di Fisica Nucleare (INFN),
the French Centre National de la Recherche Scientifique (CNRS) and
the Foundation for Fundamental Research on Matter supported by the Netherlands Organisation for Scientific Research,
for the construction and operation of the Virgo detector
and the creation and support  of the EGO consortium.
The authors also gratefully acknowledge research support from these agencies as well as by
the Council of Scientific and Industrial Research of India,
the Department of Science and Technology, India,
the Science \& Engineering Research Board (SERB), India,
the Ministry of Human Resource Development, India,
the Spanish  Agencia Estatal de Investigaci\'on,
the Vicepresid\`encia i Conselleria d'Innovaci\'o, Recerca i Turisme and the Conselleria d'Educaci\'o i Universitat del Govern de les Illes Balears,
the Conselleria d'Educaci\'o, Investigaci\'o, Cultura i Esport de la Generalitat Valenciana,
the National Science Centre of Poland,
the Swiss National Science Foundation (SNSF),
the Russian Foundation for Basic Research,
the Russian Science Foundation,
the European Commission,
the European Regional Development Funds (ERDF),
the Royal Society,
the Scottish Funding Council,
the Scottish Universities Physics Alliance,
the Hungarian Scientific Research Fund (OTKA),
the Lyon Institute of Origins (LIO),
the Paris \^{I}le-de-France Region,
the National Research, Development and Innovation Office Hungary (NKFI),
the National Research Foundation of Korea,
Industry Canada and the Province of Ontario through the Ministry of Economic Development and Innovation,
the Natural Science and Engineering Research Council Canada,
the Canadian Institute for Advanced Research,
the Brazilian Ministry of Science, Technology, Innovations, and Communications,
the International Center for Theoretical Physics South American Institute for Fundamental Research (ICTP-SAIFR),
the Research Grants Council of Hong Kong,
the National Natural Science Foundation of China (NSFC),
the Leverhulme Trust,
the Research Corporation,
the Ministry of Science and Technology (MOST), Taiwan
and
the Kavli Foundation.
The authors gratefully acknowledge the support of the NSF, STFC, MPS, INFN, CNRS and the
State of Niedersachsen/Germany for provision of computational resources.
N. Weinberg was supported in part by NASA grant NNX14AB40G.


\bibliography{../refs}

\end{document}

%% file: authors.tex
\author{B.~P.~Abbott}
\affiliation{LIGO, California Institute of Technology, Pasadena, CA 91125, USA}
\author{R.~Abbott}
\affiliation{LIGO, California Institute of Technology, Pasadena, CA 91125, USA}
\author{T.~D.~Abbott}
\affiliation{Louisiana State University, Baton Rouge, LA 70803, USA}
\author{F.~Acernese}
\affiliation{Universit\`a di Salerno, Fisciano, I-84084 Salerno, Italy}
\affiliation{INFN, Sezione di Napoli, Complesso Universitario di Monte S.Angelo, I-80126 Napoli, Italy}
\author{K.~Ackley}
\affiliation{OzGrav, School of Physics \& Astronomy, Monash University, Clayton 3800, Victoria, Australia}
\author{C.~Adams}
\affiliation{LIGO Livingston Observatory, Livingston, LA 70754, USA}
\author{T.~Adams}
\affiliation{Laboratoire d'Annecy de Physique des Particules (LAPP), Univ. Grenoble Alpes, Universit\'e Savoie Mont Blanc, CNRS/IN2P3, F-74941 Annecy, France}
\author{P.~Addesso}
\affiliation{University of Sannio at Benevento, I-82100 Benevento, Italy and INFN, Sezione di Napoli, I-80100 Napoli, Italy}
\author{R.~X.~Adhikari}
\affiliation{LIGO, California Institute of Technology, Pasadena, CA 91125, USA}
\author{V.~B.~Adya}
\affiliation{Max Planck Institute for Gravitational Physics (Albert Einstein Institute), D-30167 Hannover, Germany}
\affiliation{Leibniz Universit\"at Hannover, D-30167 Hannover, Germany}
% ...

\collaboration{The LIGO Scientific Collaboration and the Virgo Collaboration}

\author{N.~N.~Weinberg}
\affiliation{Department of Physics, and Kavli Institute for Astrophysics and Space Research, Massachusetts Institute of Technology, \\Cambridge, MA 02139, USA}